\documentclass[conference,letterpaper,twoside]{IEEEtran}

\usepackage{graphicx,epic,eepic,epsfig,amsmath,latexsym,amssymb,verbatim,color,bbm}

\usepackage{theorem}
\newtheorem{definition}{Definition}

\newtheorem{lemma}[definition]{Lemma}

\newtheorem{corollary}[definition]{Corollary}

\def\squareforqed{\hbox{\rlap{$\sqcap$}$\sqcup$}}
\def\qed{\ifmmode\squareforqed\else{\unskip\nobreak\hfil
\penalty50\hskip1em\null\nobreak\hfil\squareforqed
\parfillskip=0pt\finalhyphendemerits=0\endgraf}\fi}
\def\endenv{\ifmmode\;\else{\unskip\nobreak\hfil
\penalty50\hskip1em\null\nobreak\hfil\;
\parfillskip=0pt\finalhyphendemerits=0\endgraf}\fi}

\mathchardef\ordinarycolon\mathcode`\:
\mathcode`\:=\string"8000
\def\vcentcolon{\mathrel{\mathop\ordinarycolon}}
\begingroup \catcode`\:=\active
  \lowercase{\endgroup
  \let :\vcentcolon
  }

\newcommand{\nc}{\newcommand}
\nc{\rnc}{\renewcommand}
\nc{\beq}{\begin{equation}}
\nc{\eeq}{{\end{equation}}}
\nc{\beqa}{\begin{eqnarray}}
\nc{\eeqa}{\end{eqnarray}}
\nc{\lbar}[1]{\overline{#1}}
\nc{\bra}[1]{\langle#1|}
\nc{\ket}[1]{|#1\rangle}
\nc{\ketbra}[2]{|#1\rangle\!\langle#2|}
\nc{\braket}[2]{\langle#1|#2\rangle}
\nc{\proj}[1]{| #1\rangle\!\langle #1 |}
\nc{\avg}[1]{\langle#1\rangle}
\nc{\Rank}{\operatorname{Rank}}
\nc{\smfrac}[2]{\mbox{$\frac{#1}{#2}$}}
\nc{\tr}{\operatorname{Tr}}
\nc{\ox}{\otimes}
\nc{\dg}{\dagger}
\nc{\dn}{\downarrow}
\nc{\cA}{{\cal A}}
\nc{\cB}{{\cal B}}
\nc{\cC}{{\cal C}}
\nc{\cD}{{\cal D}}
\nc{\cE}{{\cal E}}
\nc{\cF}{{\cal F}}
\nc{\cG}{{\cal G}}
\nc{\cH}{{\cal H}}
\nc{\cI}{{\cal I}}
\nc{\cJ}{{\cal J}}
\nc{\cK}{{\cal K}}
\nc{\cL}{{\cal L}}
\nc{\cM}{{\cal M}}
\nc{\cN}{{\cal N}}
\nc{\cO}{{\cal O}}
\nc{\cP}{{\cal P}}
\nc{\cQ}{{\cal Q}}
\nc{\cR}{{\cal R}}
\nc{\cS}{{\cal S}}
\nc{\cT}{{\cal T}}
\nc{\cU}{{\cal U}}
\nc{\cW}{{\cal W}}
\nc{\cX}{{\cal X}}
\nc{\cY}{{\cal Y}}
\nc{\cZ}{{\cal Z}}
\nc{\csupp}{{\operatorname{csupp}}}
\nc{\qsupp}{{\operatorname{qsupp}}}
\nc{\var}{{\operatorname{var}}}
\nc{\Var}{{\operatorname{Var}}}
\nc{\rar}{\rightarrow}
\nc{\lrar}{\longrightarrow}
\nc{\polylog}{{\operatorname{polylog}}}
\nc{\1}{{\mathbbm{1}}}
\nc{\wt}{{\operatorname{wt}}}
\nc{\av}[1]{{\left\langle {#1} \right\rangle}}

\nc{\RR}{{{\mathbb R}}}
\nc{\CC}{{{\mathbb C}}}
\nc{\DD}{{{\mathbb D}}}
\nc{\FF}{{{\mathbb F}}}
\nc{\NN}{{{\mathbb N}}}
\nc{\ZZ}{{{\mathbb Z}}}
\nc{\PP}{{{\mathbb P}}}
\nc{\QQ}{{{\mathbb Q}}}
\nc{\UU}{{{\mathbb U}}}
\nc{\EE}{{{\mathbb E}}}
\nc{\id}{{\operatorname{id}}}

\nc{\CHSH}{{\operatorname{CHSH}}}

\nc{\be}{\begin{equation}}
\nc{\ee}{{\end{equation}}}
\nc{\bea}{\begin{eqnarray}}
\nc{\eea}{\end{eqnarray}}
\nc{\<}{\langle}
\rnc{\>}{\rangle}
\nc{\Hom}[2]{\mbox{Hom}(\CC^{#1},\CC^{#2})}
\nc{\rU}{\mbox{U}}

\nc{\ob}[1]{#1}

\nc{\SEP}{{\text{SEP}}}
\nc{\NS}{{\text{NS}}}
\nc{\LOCC}{{\text{LOCC}}}
\nc{\PPT}{{\text{PPT}}}
\nc{\EXT}{{\text{EXT}}}
\nc{\Sym}{{\operatorname{Sym}}}

\nc{\ERLO}{{E_{\text{r,LO}}}}
\nc{\ERLOCC}{{E_{\text{r,LOCC}}}}
\nc{\ERPPT}{{E_{\text{r,PPT}}}}
\nc{\ERLOCCinfty}{{E^{\infty}_{\text{r,LOCC}}}}
\nc{\Aram}{{\operatorname{\sf A}}}

\begin{document}

\title{An Alphabet-Size Bound for the \protect\\ Information Bottleneck Function}

\date{22 November 2019}

\author{%
  \IEEEauthorblockN{Christoph Hirche\IEEEauthorrefmark{1}}
  \IEEEauthorblockA{\IEEEauthorrefmark{1}%
                    QMATH\\
                    Department of Mathematical Sciences\\
                    University of Copenhagen\\
                    Universitetsparken 5\\
                    2100 Copenhagen, Denmark\\
                    Email: christoph.hirche@gmail.com}
  \and
  \IEEEauthorblockN{Andreas Winter\IEEEauthorrefmark{2}\IEEEauthorrefmark{3}}
  \IEEEauthorblockA{\IEEEauthorrefmark{2}%
                    Grup d'Informaci\'{o} Qu\`{a}ntica,
                    Departament de F\'{\i}sica\\
                    Universitat Aut\`{o}noma de Barcelona\\
                    08193 Bellaterra (Barcelona), Spain \vspace{1mm}}
  \IEEEauthorblockA{\IEEEauthorrefmark{3}%
                    Instituci\'o Catalana de
                    la Recerca i d'Estudis Avan\c{c}ats\\ 
                    Pg.~Lluis Companys, 23, 
                    08010 Barcelona, Spain\\
                    Email: andreas.winter@uab.cat}
}

\maketitle

\begin{abstract}
The information bottleneck function gives a measure of optimal preservation of 
correlation between some random variable $X$ and some side information $Y$ while 
compressing $X$ into a new random variable $W$ with bounded remaining correlation 
to $X$. As such, the information bottleneck has found many natural applications 
in machine learning, coding and video compression. The main objective in order 
to calculate the information bottleneck is to find the optimal representation 
on $W$. This could in principle be arbitrarily complicated, but fortunately it 
is known that the cardinality of $W$ can be restricted as $|\cW|\leq|\cX|+1$ which 
makes the calculation possible for finite $|\cX|$. Now, for many practical 
applications, e.g. in machine learning, $X$ represents a potentially very 
large data space, while $Y$ is from a comparably small set of labels. This 
raises the question whether the known cardinality bound can be improved in such 
situations. We show that the information bottleneck function can always be 
approximated up to an error $\delta(\epsilon,|\cY|)$ with a cardinality 
$|\cW| \leq f(\epsilon,|\cY|)$, for explicitly given functions $\delta$ and $f$ 
of an approximation parameter $\epsilon>0$ and the cardinality of $\cY$.
 
Finally, we generalize the known cardinality bounds to the case were some of 
the random variables represent quantum information. 
\end{abstract}

\section{Introduction}
\label{sec:intro}
Given a joint probability distribution $P_{XY}$, where $X$ is often interpreted 
as some kind of data and $Y$ as some side information (e.g.~labels in supervised 
machine learning), the information bottleneck function~\cite{tishbyIBM} quantifies 
the optimal compression of $X$ into $W$ below a certain threshold while preserving 
the maximal amount of correlation with the side information. This is closely 
related to the question of finding approximate sufficient statistics for $P_{XY}$ 
and has therefore found a multitude of applications such as e.g.~in investigating 
deep neural networks~\cite{tishby2015deep, shwartz2017opening}, video processing~\cite{hsu2006video}, clustering~\cite{slonim2000doc} and polar coding~\cite{StarkPolar}.
Formally, the information
bottleneck function of a given joint distribution $P_{XY}$ of two
random variables $X$ and $Y$ is defined as~\cite{tishbyIBM}
\begin{align}
  I_{XY}(R) &:= \max I(Y:W) 
                \text{ s.t. } Y-X-W \text{ Markov chain, } \nonumber \\
            &\phantom{:= \max I(Y:W) \text{ s.t.. }} I(X:W) \leq R.
\end{align}
Here, $X$ is considered the ``data'' and $Y$ the ``classifier'', which is
assumed to capture the objective truth about each data point, or at least
reflect it on a training set.

It is known, and not difficult to prove using Caratheodory's theorem,
that the above maximum is attained with $|\cW| \leq |\cX|+1$, 
which makes the information bottleneck function at least in principle 
computable. 
By Caratheodory's theorem we refer to the fundamental geometric fact 
that if a point in an $n$-dimensional real affine space, w.l.o.g. $\RR^n$, 
is a convex combination of any set $\cE$, then it can be written as 
a convex combination of a subset of cardinality $n+1$ \cite{Rockafeller}.
(Some earlier work states the bound $|\cW| \leq |\cX|+2$, however directly 
using the results in~\cite{WW1975} gives the slightly better bound, see also~\cite{hsu2018}.) 
However, in the purported machine learning applications, $X$ takes 
values in an enormously large alphabet, e.g. digital pictures of
a certain format, which would require optimisation over channels $P_{W|X}$
with potentially megabyte-sized input and output, which is clearly 
infeasible. On the other hand, $Y$ is by definition from a small set,
for example letters or numbers in the task of recognizing written text.

It was found that in some cases deep neural networks approximate the information bottleneck 
function well, despite their output alphabet seemingly being too small, or
at least much smaller than the above guarantee~\cite{shwartz2017opening}. 
This raises the question whether the information bottleneck can generally be well 
approximated with a smaller alphabet size $|\cW|$, and in particular one with bounded 
cardinality depending on $Y$ rather than on $X$. In the present paper we answer 
this question affirmatively, by 
\begin{enumerate}
  \item providing an explicit method that reduces the cardinality of $X$, by compressing 
        it into a random variable $X'$ whose range is determined solely by $|\cY|$ and 
        an error parameter $\epsilon>0$, that allows to approximately recover the 
        statistics of $X$ conditioned on $Y$ from $X'$; 
  \item showing how this implies that the information bottleneck function of $P_{XY}$
        is well-approximated by that of the compressed $P_{X'Y}$; and 
  \item showing that the same tools imply that evaluating the information bottleneck 
        function on the original $P_{XY}$, but with bounded cardinality 
        $|\cW| \leq f(\epsilon, |\cY|)\equiv N$, also gives a good approximation of 
        the unbounded case.  
\end{enumerate} 

Besides the usual information bottleneck function, recently also a version based on quantum 
information was considered~\cite{Grimsmo16,salek2017,DHW2019}. Dimension bounds on 
auxiliary quantum registers in quantum information theory are generally a much harder 
problem since no good tools for investigating them are known. 
In the final section of this work, we will generalize the previously known 
bound and the ones developed in this work to particular intermediate settings where some 
random variables are classical and others are quantum.

\section{Approximation via recovery}
\label{sec:recovery}
The following lemma shows how to obtain an approximately sufficient statistics 
for $X$ with respect to $Y$, with an alphabet size that depends only on $|\cY|$
and on the accuracy of the approximation. The main tool is the existence of a recovery 
map that can approximately revert the compression. 

\begin{lemma}
  \label{lemma:approximate}
  Given a joint distribution $P_{XY}$ of two random variables $X$ and $Y$,
  and assuming that there exist $N$ probability distributions $Q_1,\ldots,Q_N$ 
  on $\cY$, and a function $f:\cX\longrightarrow[N]$ with the property that 
\begin{align}\label{eqn:boundeps}
    \forall x \quad \frac12 \| P_{Y|X=x} - Q_{f(x)} \|_1 \leq \epsilon,
\end{align}
for some $\epsilon>0$. 
  Then there exists a recovery channel $S:[N]\longrightarrow\cX$ 
  such that the Markov chain $Y - X - X' - \widehat{X}$ defined by
  $X'=f(X)$ and $P_{\widehat{X}|X'}=S$ satisfies
  $P_X = P_{\widehat{X}}$ and
  $\frac12 \| P_{XY} - P_{\widehat{X}Y} \|_1 \leq \epsilon' = 2\epsilon$.
\end{lemma}
\begin{proof}
The function $f$ defines a partition of $\cX = \dot{\bigcup}_{i} \cX_i$
into the pre-images $\cX_i = f^{-1}(i)$. Define 
\[
  S(x|i) = \begin{cases}
             \frac{1}{P_X(\cX_i)}P_X(x) & \text{ if } x\in \cX_i, \\
             0                          & \text{ otherwise}.
           \end{cases}
\]
It can easily be checked that this leads to the desired results by calculating 
$P_{\widehat{X}Y}$ and then bounding $\frac12 \| P_{XY} - P_{\widehat{X}Y} \|_1$ 
using triangle inequality and twice the assumption in Eq.~\ref{eqn:boundeps}.
It actually furthermore has the property
that it gives a Markov chain $Y - X - X' - \widehat{X} - X'$, since
$X'=f(X)=f(\widehat{X})$ by construction.
\end{proof}

\medskip
The idea, in any case, is that $f$ identifies different $x$ that have 
almost the same correlation with $Y$, as expressed by the conditional
distribution; $S$ recovers $X$ as far as its correlation with $Y$ is concerned.
Thus, $X'$ is an approximately sufficient statistics for $X$.

\medskip
Now the question is: How large does $N$ need to be? Surely not 
larger than $|\cX|$, but we assume that to
be potentially unbounded. In the worst case, we need to choose an $\epsilon$-net of
the probability simplex $\cP(\cY)$ of all probability distributions on
$\cY$ with respect to the total variational distance, which results in
$N \leq \left(\frac{c}{\epsilon}\right)^{|\cY|}$, with some universal
constant $c>0$.

To be more precise, $N$ is the minimum cardinality of an
$\epsilon$-net, and a standard upper bound is given by the corresponding covering 
number. A standard estimate for the covering number of $K\subset \RR^n$ by a 
convex and symmetric $D\subset \RR^n$ can be attained via the volume of the involved sets
\begin{align}
  \frac{vol(K)}{vol(D)} \leq N \leq \frac{vol(K+\frac{1}{2}D)}{vol(\frac12 D)}.
\end{align}
A probability simplex $\Delta_\cY$ can be understood as a subset of the unit 
sphere in $\RR^{|\cY|}$ which again is the boundary of the corresponding unit ball. 
Using the above volume bound for $\epsilon$-nets on the unit ball by $\epsilon$-balls, we get that
\begin{align}
N_\Delta \leq \left( \frac{2}{\epsilon} + 1 \right)^{|\cY|} \leq \left( \frac{3}{\epsilon} \right)^{|\cY|} .
\end{align}
This is in fact independent of the metric used, as long as the unit 
ball and the $\epsilon$-ball are defined via the same metric.
However, using somewhat more involved ideas it is possible to determine 
a bound on the covering number of the unit sphere more directly 
(see~\cite[Lemma~5.3]{AliceBobBanach}), leading to
\begin{align}
  N_\Delta \leq \left( \frac{2}{\epsilon} \right)^{|\cY|}.
\end{align} 
We will however get back to using the simpler estimates when discussing the classical-quantum case.

\section{Application to approximating the \protect\\ information bottleneck function.}
\label{sec:IB}
To apply the results from the last section to find the desired approximation 
results, we first prove another lemma, exploiting the Markov chain emerging
from Lemma \ref{lemma:approximate}.

\begin{lemma}
  \label{lemma:IB-Markov}
  Let $Y - X - \widetilde{X}$ be a Markov chain. Then the IB function of 
  $P_{XY}$ dominates the IB function of $P_{\widetilde{X}Y}$ pointwise:
  \[
    I_{XY}(R) \geq I_{\widetilde{X}Y}(R) \quad \forall R.
  \]
\end{lemma}
\begin{proof}
This follows easily from the definition: recall that 
\begin{equation}\begin{split}
  I_{\widetilde{X}Y}(R) &= \max I(Y:W) \text{ with Markov chain} \nonumber\\
                        &\phantom{=========} 
                                       Y-X-\widetilde{X}-W,\, I(\widetilde{X}:W) \leq R.
\end{split}\end{equation}
Hence, given any $W$ eligible for $\widetilde{X}Y$, by data processing inequality
we have $I(X:W)\leq R$, and so $W$ is also eligible for $XY$. Since we are
optimising the same objective function, the claim follows.
\end{proof}

\medskip
The lemma implies that for the Markov chain $Y - X - X' - \widehat{X} - X'$ resulting from 
Lemma \ref{lemma:approximate}, we have
\begin{equation*}
  I_{XY}(R) \geq I_{X'Y}(R) \geq I_{\widehat{X}Y}(R) \geq I_{X'Y}(R),
\end{equation*}
which implies
\begin{equation}
  \label{eq:IB-suff}
  I_{XY}(R) \geq I_{X'Y}(R) = I_{\widehat{X}Y}(R).
\end{equation}

Now we are ready for our main result, which bounds the gap in the latter inequality:
\begin{corollary}
  \label{cor:approximate-IB}
  Under the assumptions of Lemma \ref{lemma:approximate}, 
  \[
    I_{X'Y}(R) \leq I_{XY}(R) \leq I_{X'Y}(R) + \delta(\epsilon),
  \]
  where $\delta(\epsilon, |\cY|) := \epsilon'\log|\cY| + (1\!+\!\epsilon')h\!\left(\!\frac{\epsilon'}{1+\epsilon'}\!\right)$.
\end{corollary}
\begin{proof}
The left hand side inequality has been shown in Lemma \ref{lemma:IB-Markov},
concretely the inequality in Eq.~(\ref{eq:IB-suff}).
For the right hand side bound, we use the equality in Eq.~(\ref{eq:IB-suff}),
and consider a channel $P_{W|X}:\cX\longrightarrow\cW$.
Define Markov chains $Y-X-W$ and $Y-\widehat{X}-\widehat{W}$ by letting
\[
  P_{YXW} = P_{XY}P_{W|X}, \quad
  P_{Y\widehat{X}\widehat{W}} = P_{\widehat{X}Y}P_{\widehat{W}|\widehat{X}},
\]
where we set $P_{\widehat{W}|\widehat{X}} = P_{W|X}$.
Then we have $I(X:W)=I(\widehat{X}:\widehat{W})$, since $P_X = P_{\widehat{X}}$
and hence $P_{WX} = P_{\widehat{W}\widehat{X}}$. That is, $W$ is eligible for $XY$
if and only if $\widehat{W}$ is eligible for $\widehat{X}Y$. 

On the other hand, 
\[
  \bigl| I(Y:W)-I(Y:\widehat{W}) \bigr| = \bigl| H(Y|W)-H(Y|\widehat{W}) \bigr| \leq \delta(\epsilon, |\cY|),
\]
where we have used the Alicki-Fannes continuity bound for the conditional entropy
in the form of \cite{S-conti}.
\end{proof}

\medskip
For Lemma~\ref{lemma:approximate} to be applicable, we can use an $\epsilon$-net 
for the probability distributions on $\cY$, 
resulting in $N\leq \left(\frac{2}{\epsilon}\right)^{|\cY|}$. 
This means of course, that to approximate $I_{XY}$ we might as well 
compute $I_{X'Y}$, and the latter can attain its optimal value with an alphabet
size $|\cW| \leq |\cX'|+1 = N+1 \leq \left(\frac{2}{\epsilon}\right)^{|\cY|} + 1$.

In practice, utilizing Corollary~\ref{cor:approximate-IB} will require some simple 
preprocessing on the used probability distribution associated with the considered 
dataset in order to calculate the simplified bottleneck function. In some cases, 
one might want to directly use the original distribution, but still restrict $|\cW|$. 
This should in particular be useful when $\cX$ is very large. In the following we 
denote by $I_{XY}(R,N)$ the bottleneck function $I_{XY}(R)$ with the additional 
constraint that $|\cW|\leq N$. 

\begin{corollary}
  \label{cor:approximate-IB-2}
  Under the assumptions of Lemma \ref{lemma:approximate}, 
  \[
    I_{XY}(R,N) \leq I_{XY}(R) \leq I_{XY}(R,N) + \delta(\epsilon, |\cY|),
  \]
  where $\delta(\epsilon, |\cY|) := \epsilon'\log|\cY| + (1\!+\!\epsilon')h\!\left(\!\frac{\epsilon'}{1+\epsilon'}\!\right)$ 
  and $N\leq \left(\frac{2}{\epsilon}\right)^{|\cY|}$. 
\end{corollary}
\par\noindent\hspace{2em}{\itshape Proof: }
Using the definitions in Lemma~\ref{lemma:approximate} and applying it to the 
joint distribution $P_{YW}$ resulting from the optimal channel $P_{W|X}$, we get 
the following Markov chain, $Y- X - W - W' -\widehat{W}$. Here, $W'$ is constructed 
via an $\epsilon$-net on $Y$ and therefore requires at most 
$N\leq \left(\frac{2}{\epsilon}\right)^{|\cY|}$. 
Furthermore, we have $ \frac12 \| P_{YW} - P_{Y\widehat{W}} \|_1 \leq 2\epsilon$. 
It is immediately clear that restricting the maximization to a smaller cardinality can 
not increase the result and therefore the left hand side follows directly. 

Given that $I(Y:W)$ is the optimal value for $I_{XY}(R)$, we can use data processing 
and a continuity bound to get
\begin{equation}
  I(Y \!:\! W)  \geq \! I(Y \!:\! W') 
          \geq \! I(Y \!:\! \widehat{W})
          \geq \! I(Y \!:\! W) \!-\! \delta(\epsilon, |\cY|). 
\end{equation}
From data processing we also immediately get that  
\begin{align}
  I(X:W) \geq I(X:W'), 
\end{align}
which means that for a given rate $R$, if $P_{W|X}$ is an eligible map, then also $P_{W'|X}$ is. 
Therefore, we observe that the restricted bottleneck function deviates from the 
unrestricted one by at most $\delta(\epsilon)$:
\begin{equation*}
  \hspace{0.8cm} I_{XY}(R) - \delta(\epsilon, |\cY|) \leq I(Y:W') \leq I_{XY}(R,N). \hspace{0.9cm} \IEEEQED
\end{equation*}

While this corollary gives a slightly more direct approach to the problem, it has the 
downside that, a priori, we do not know anything about the possible distributions 
$P_{Y|w=w}$. Therefore, in general an $\epsilon$-net on $\cY$ is needed.

\subsection{Examples}
For many applications, such as machine learning, the involved probability distribution 
$P_{XY}$ often stems from a dataset of objects with assigned labels.  These datasets 
are then used as training sets e.g. towards a classification task. Generally, if we 
have a multi-class classification problem every object is assigned exactly one label. 
The number of labels available determines $|\cY|$. Now, it might be the case that some 
objects appear multiple times in the dataset but with different labels; let us call such 
objects \textit{ambiguous}. However, in the framework of Corollary~\ref{cor:approximate-IB}, 
as long as the total number of such objects in the dataset is not too large, only a very 
restricted number of distributions $P_{Y|X=x}$ can appear and we can even achieve $\epsilon=0$. 
 
A particular case that is often encountered is that of deterministic datasets, i.e. $Y=f(X)$ 
for some single-valued function $f$, meaning that there are no ambiguous objects. This includes 
many commonly used datasets such as MNIST and CIFAR. It is known that in these cases the 
information bottleneck function takes a very simple form~\cite{kolchinsky2018caveats}. 
This is also the case when considering the conditions for our Lemma~\ref{lemma:approximate} 
since the possible probability distributions are limited to $P_{Y|X=x} = \delta_{y,f(x)}$, 
and therefore $N = |\cY|$ is sufficient to even achieve $\epsilon=0$. 
 
For a small amount of ambiguous objects, let's say $a$ of them, one can still achieve $\epsilon=0$ 
with $N = |\cY|+a$. However, for large datasets, $a$ might quickly become unpractical and 
the approximating method via $\epsilon$-nets becomes much more viable. 
 
One could also consider a different class of datasets, namely those for multi-label datasets 
(every object can be assigned multiple labels). Let's assume that the dataset does not 
include any ambiguous objects. Now, with a little combinatorics one sees that we 
can achieve $\epsilon=0$ with $N = 2^{|\cY|}$ if the number of labels per objects is only 
restricted by the total number of labels $|\cY|$. If every object is assigned exactly $k$ 
labels we can achieve the same with $N = \dbinom{|\cY|}{k}$ and with a maximum of $k$ 
labels per object, $N = \sum_{i=0}^k \dbinom{|\cY|}{i}$ suffices. While these quantities 
are not necessarily small, they are still significantly better than the $\epsilon$-net 
approximation method.

\section{Classical-quantum setting}
\label{sec:cq}
In quantum information theory the role of probability distributions is taken by 
quantum states, i.e. positive semi-definite hermitian matrices with trace one. 
Instead of maps between probabilities, these states are transformed by completely 
positive and trace preserving maps called quantum channels. For a complete 
introduction we refer the reader to the literature, e.g.~\cite{wilde2013quantum}. 

In the traditional (classical) setting the main tool to prove alphabet size bounds 
is Caratheodory's theorem. One crucial property that allows us to apply 
Caratheodory's theorem is that every conditional entropy for a classical 
probability distribution $p_{XY}$ can be written as 
\begin{align}\label{Eq:sumprop}
  H(X|W) = \sum_w p(w) H(X|W=w). 
\end{align}
This might seem like a trivial observation, however it should be noted that when 
we move from classical probabilities to quantum states, in general such an equality 
does not exist, leading to major complications in quantum information 
theory~\cite{beigi2013dimension, hirche18}. In the following we will discuss an 
intermediate setting where the random variables $X, Y$ are quantum and only $W$ 
is restricted to be classical. In this setting, the generalization of Eq.~(\ref{Eq:sumprop}) 
still holds and we will be able to apply Caratheodory's theorem as well as the new 
framework developed in the first part of this work.  

Let us briefly recall the definition of the quantum information bottleneck. 
In~\cite{Grimsmo16} it was defined as the natural generalization of the classical case as
\begin{align}
I^q_{XY}(R) &:= \sup_{\cN^{X\to W}\atop{I(X';W)_{\tilde{\tau}} \leq R}} I(Y;W)_\sigma,
\end{align}
with
\begin{align}\label{ttau}
  {\tilde{\tau}}_{X'W} &:= ({\rm{id}}_{X'} \otimes \cN^{X\to W})\tau_{X'X},
\end{align}
where $\tau_{X'X}$ is a purification of $\rho_X$, and 
\begin{equation}
  \sigma_{WY} := ( \cN^{X\to W} \otimes {\rm{id}}_{Y} )\rho_{XY}. 
\end{equation}
Hence $\rho_X = \tr_{X'}\tau_{X'X} = \tr_Y \rho_{XY}$.
This formula unfortunately proved difficult to handle analytically. It was however 
shown in~\cite{DHW2019} that the following formulation is equivalent to the one above:
\begin{align}
  I^q_{XY}(R) &= \sup_{\cN^{X\to W}\atop{I(YR;W)_\sigma \leq R}} I(Y;W)_\sigma, 
\end{align}
where  $\sigma_{WYR} := (\cN^{X \to W} \otimes \id_{YR}) \psi_{XYR}$ and $\psi_{XYR}$ 
a purification of $\rho_{XY}$. 

Defining the quantum information bottleneck via purifications is in general necessary, 
due to the well known fact that quantum information cannot be copied. In the 
intermediate case where only $Y$ is quantum and $X, W$ are classical, this reduces 
to a definition that more closely resembles the fully classical case:
\begin{align}
  I^{cq}_{XY}(R) := \sup_{\cN^{X\to W}\atop{I(X;W)_\sigma \leq R}} I(Y;W)_\sigma,
\end{align}
where purifying the state $\rho_{XY}$ becomes equivalent to copying the classical $X$. 

First, we will argue that the classical bound proven by using Caratheodory's theorem 
can be generalized to the classical-quantum setting where $X$ and $Y$ are quantum, and 
$W$ is classical, leading to the following lemma. 
\begin{lemma}
  For $X$ and $Y$ quantum, and $W$ classical, an optimal solution for the quantum 
  information bottleneck can be achieved with $|\cW| \leq |\cY|^2|\cR|^2+1$. 
\end{lemma} 
\begin{proof}
The proof idea in the classical case comes originally from~\cite{WW1975}, see 
also~\cite{hsu2018} for an application to generalized classical information bottleneck 
functions. 
Starting from a state $\psi_{XYR}$ we consider the setting where the channel $\cN$ 
has a quantum input but a classical output. The resulting state $\sigma_{WYR}$ 
will then have the following classical-quantum form
\begin{align}
  \sigma_{WYR} = \sum_w p(w) |w\rangle\langle w| \otimes \rho_{YR}^w, 
\end{align}
with $\{|w\rangle\}_w$ some orthonormal basis on $W$. Now, optimizing over the 
channel is equivalent to optimizing over all $\{ p(w), |w\rangle \}$. 
The feasible solutions for the quantum information bottleneck in this setting 
are now completely determined by the set
\begin{align}
\cC&(\Psi_{XYR}) \nonumber\\
   &=\left\{\! \bigl(\rho_{YR},  I(YR \!:\! W), I(Y \!:\! W)  \bigr) \!:\! 
                \sum_w p(w) \rho_{YR}^w \!=\! \rho_{YR} \!\right\} \nonumber\\
   &=\left\{\! \left(\! \rho_{YR}, \sum_w p(w) [ H(YR) \!-\! H(YR | W\!=\!w)],\right.\right. \\
   &\phantom{===}
    \left.\left. \sum_w p(w) [ H(Y) \!-\! H(Y | W\!=\!w) ] \!\right) \!\!:\! 
                              \sum_w \! p(w) \rho_{Y\!R}^w \!=\! \rho_{Y\!R} \!\right\} \!\!. \nonumber
\end{align}
Following the argument in~\cite{WW1975}, $\cC(\Psi_{XYR})$ is a convex and compact set. 
Also, the state $\rho_{YR}$ can be interpreted as living on a $|\cY|^2|\cR|^2$ real 
vector space, therefore a direct application of Caratheodory's theorem proves the above lemma. 
\end{proof}

\medskip
Note that $\psi$ can always be chosen to be a minimal purification of $\rho$ with 
$|\cR| \leq |\cX||\cY|$, and therefore we have $|\cW| \leq |\cY|^4 |\cX|^2+1$. 

One can also consider a setting in between the fully classical case and the 
one considered before, that leads to a bound closer to what is familiar in the classical case. 

\begin{lemma}
For $Y$ quantum, but $X$ and $W$ classical, an optimal solution for the quantum information 
bottleneck can be achieved with $|\cW| \leq |\cX|+1$. 
\end{lemma} 
The proof is almost identical to that of the fully classical case and omitted here for brevity. 

In this second intermediate scenario we can also find a classical-quantum version of 
the cardinality bounds achieved with the recoverability strategy. Therefore, we first 
need a generalization of Lemma~\ref{lemma:approximate}.

\begin{lemma}
  \label{lemma:approximateQ}
  Given a classical-quantum state 
  \begin{align}
    \rho_{XY} = \sum_x p(x) |x\rangle\langle x | \otimes \rho_Y^x, 
  \end{align}
  and assume that there exist $N$ quantum states $\sigma_Y^1,\ldots,\sigma_Y^N$  
  and a function $f:\cX\longrightarrow[N]$ with the property that 
  \begin{align}
    \forall x \quad \frac12 \| \rho_Y^x - \sigma_Y^{f(x)} \|_1 \leq \epsilon,
  \end{align}
  for given $\epsilon>0$. 
  Then there exists a recovery channel $S:[N]\longrightarrow\cX$ 
  such that the Markov chain $Y - X - X' - \widehat{X}$ defined by
  $X'=f(X)$ and $P_{\widehat{X}|X'}=S$ satisfies
  $P_X = P_{\widehat{X}}$ and
  $\frac12 \| \rho_{XY} - \rho_{\widehat{X}Y} \|_1 \leq \epsilon' = 2\epsilon$.
\end{lemma}

\begin{proof}
The proof is similar to the completely classical case. Again, the function $f$ defines 
a partition of $\cX = \dot{\bigcup}_{i} \cX_i$
into the pre-images $\cX_i = f^{-1}(i)$. Now, it can be shown that the function
\[
  S(x|i) = \begin{cases}
             \frac{1}{P_X(\cX_i)}P_X(x) & \text{ if } x\in \cX_i, \\
             0                          & \text{ otherwise},
           \end{cases}
\]
achieves the desired result. The above strategy again has the property
that it gives a Markov chain $Y - X - X' - \widehat{X} - X'$, since
$X'=f(X)=f(\widehat{X})$ by construction.
\end{proof}

\medskip
Again, we shall ask what is the necessary $N$ for the assumptions to hold in general. 
Following the classical argument it is always possible to satisfy the assumptions 
by choosing an $\epsilon$-net, but now on the set of quantum states. Pure quantum 
states correspond to the unit sphere in $\CC^{|\cY|} \cong \RR^{2|\cY|}$. Mixed quantum 
states can be represented by pure states on a purifying system $\cY \otimes \cY'$,
whose Hilbert space dimension is $|\cY|^2$. 
We can now use the previously presented volume bound to get 
\begin{align}
  N_Q \leq \left( \frac{2}{\epsilon} + 1 \right)^{2|\cY|^2} \leq \left( \frac{3}{\epsilon} \right)^{2|\cY|^2} .
\end{align}

Using the result in Lemma~\ref{lemma:approximateQ}, we immediately get the following generalizations.

\begin{lemma}
  \label{lemma:IB-MarkovQ}
  Let $Y - X - \widetilde{X}$ be a Markov chain. Then the IB function of 
  $\rho_{XY}$ dominates the IB function of $\rho_{\widetilde{X}Y}$ pointwise:
  \[
    I^{cq}_{XY}(R) \geq I^{cq}_{\widetilde{X}Y}(R) \quad \forall R.
  \]
\end{lemma}

\begin{corollary}
  \label{cor:approximate-IBQ}
  Under the assumptions of Lemma \ref{lemma:approximateQ}, 
  \[
    I^{cq}_{X'Y}(R) \leq I^{cq}_{XY}(R) \leq I^{cq}_{X'Y}(R) + \delta(\epsilon, |\cY|),
  \]
  where $\delta(\epsilon, |\cY|) := \epsilon'\log|\cY| + (1\!+\!\epsilon')h\!\left(\!\frac{\epsilon'}{1+\epsilon'}\!\right)$.
\end{corollary}

\begin{corollary}
  \label{cor:approximate-IB-2Q}
  Under the assumptions of Lemma \ref{lemma:approximateQ}, 
  \[
    I^{cq}_{XY}(R,N) \leq I^{cq}_{XY}(R) \leq I^{cq}_{XY}(R,N) + \delta(\epsilon, |\cY|),
  \]
  where $\delta(\epsilon, |\cY|)$ is as before
  and $N\leq \left(\frac{3}{\epsilon}\right)^{2|\cY|^2}$. 
\end{corollary}

\medskip
The proofs of the last three results are very similar to those in the fully classical 
case and therefore omitted here for the sake of brevity. 

\section{Conclusion}
In the present paper we addressed the question whether the information bottleneck 
function can be well-approximated with an auxiliary random variable whose alphabet size 
$|\cW|$ only depends on $|\cY|$ and the allowed error. We proved that this is 
indeed the case, using an approach based on approximate sufficiency, which 
itself relates to approximate Markov chains through (approximate) recovery maps. 
We showed that for many practical problems this approach drastically 
reduces the cardinality needed and therefore the computational complexity of evaluating 
the bottleneck function. In the second part, we discussed extensions of these and other 
known bounds to settings with quantum information and showed that in some special cases 
dimension bounds exist and make it possible to evaluate the quantum information bottleneck. 

Furthermore our results, in particular Lemma~\ref{lemma:approximate}, could provide a 
possible road towards classifying problems which are suitable to be solved by neural 
networks. The general approach of this work could also lead to new cardinality bounds 
for other information theoretic quantities. 

An important question left open is to find dimension bounds for the fully quantum case, 
a setting where the usual tools like Caratheodory's theorem do not seem to be applicable. 
We 
discuss the application of our new recoverability approach to the fully quantum 
setting in~\cite{CHWsoon}.

\bigskip\noindent
\textbf{Acknowledgments.} 
The authors thank Axel Foley for discussions on information recovery. 
CH was supported by the VILLUM FONDEN via
the QMATH Centre of Excellence (Grant no. 10059).
AW was supported by the Spanish MINECO, projects
FIS2016-86681-P, with the support of FEDER funds; and
by the Generalitat de Catalunya, CIRIT project 2014-SGR-966.

\bibliographystyle{IEEEtran}



\end{document}